\documentclass[sigconf,nonacm]{acmart}

\setcopyright{none}
\renewcommand\footnotetextcopyrightpermission[1]{}

\AtBeginDocument{%
  }

\usepackage{booktabs}
\usepackage{amsmath}

\usepackage{placeins}
\begin{document}

\title{When2Talk: When Should a Proactive In-Car Agent Talk?}

\author{Kaiser Hamid}
\authornote{Kaiser Hamid and Peihang Li contributed equally to this work.}
\affiliation{%
  \institution{Texas Tech University}
  \city{Lubbock}
  \state{Texas}
  \country{USA}}

\author{Peihang Li}
\authornotemark[1]
\affiliation{%
  \institution{Texas Tech University}
  \city{Lubbock}
  \state{Texas}
  \country{USA}}

\author{Nade Liang}
\correspondingauthor
\affiliation{%
  \institution{Texas Tech University}
  \city{Lubbock}
  \state{Texas}
  \country{USA}}
\email{nade.liang@ttu.edu}

\renewcommand{\shortauthors}{Hamid, Li, and Liang}

\newcommand{\ET}{\textsc{ET}}
\newcommand{\CS}{\textsc{CS}}
\newcommand{\sys}{\textsc{When2Talk}}
\begin{abstract}
Proactive in-cabin agents can help passengers understand automated-vehicle (AV) behavior, but communicating every ride event may 
introduce unnecessary interruptions. We investigated how communication should adapt to event priority and passenger activity. In a 
mixed-methods within-subject study, 41 participants rode as passenger in a VR simulated fully-automated vehicle. We compared an 
event-triggered (ET) policy that communicated immediately at every event with a context-sensitive (CS) policy that selected 
\textit{Immediate}, \textit{Delayed}, or \textit{Silent} communications. CS increased communication appropriateness and substantially 
reduced perceived interruption. Perceived trust did not differ between policies, although baselines dispositional trust  differentiated 
communication preferences. Findings highlight event consequence, passenger activity, continuing information value, and confirmation 
need as key considerations for selective in-cabin communication.
\end{abstract}


\begin{teaserfigure}
    \centering
    \includegraphics[width=\textwidth]{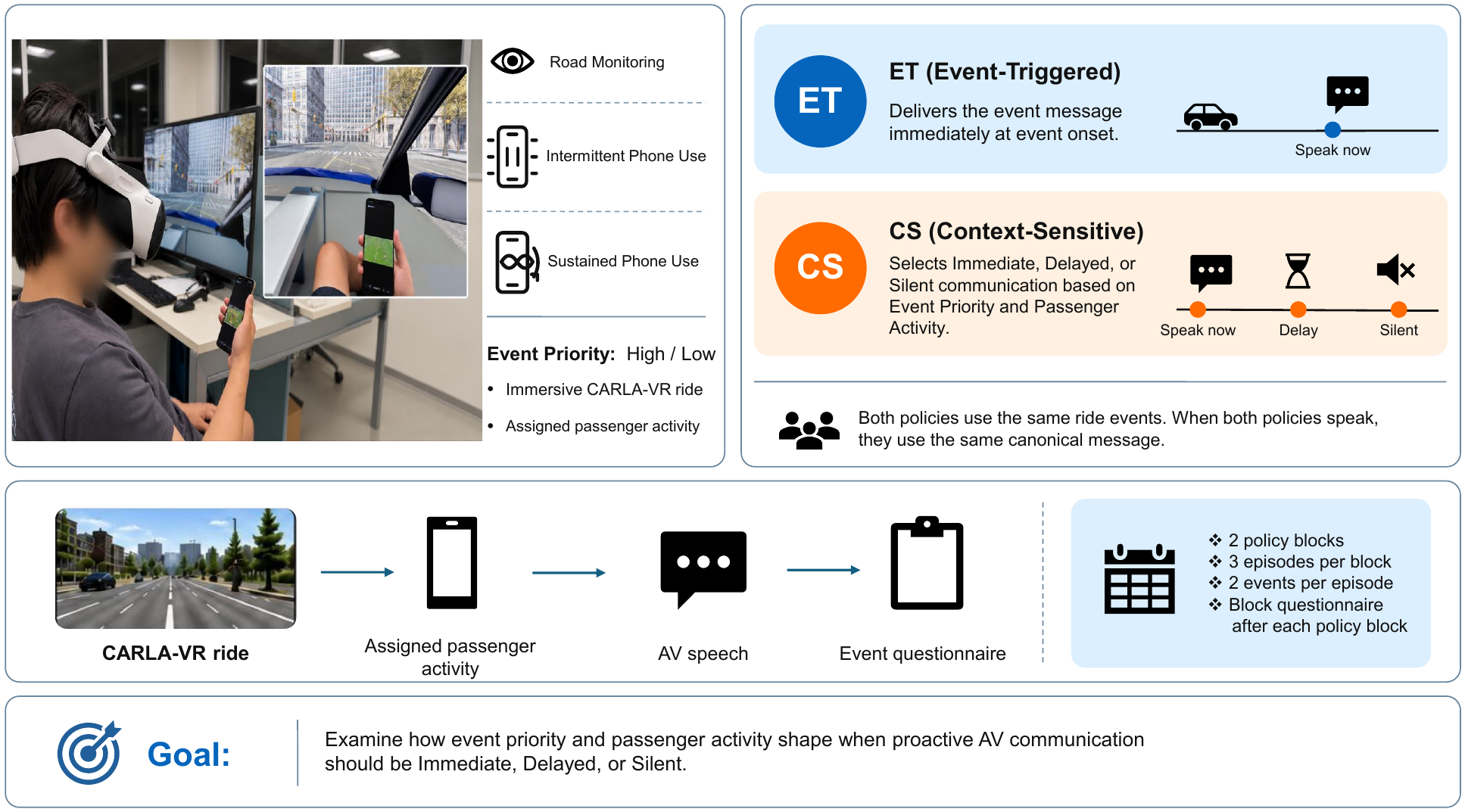}
    \caption{\textbf{When2Talk study overview.}
In a controlled CARLA/VR passenger-seat simulation, we compare event-triggered (ET) communication, which speaks immediately at every driving event, with context-sensitive (CS) communication, which considers event priority and passenger activity to speak immediately, defer the message, or remain silent.}
    \label{fig:teaser}
\end{teaserfigure}

\maketitle
\begin{center}
\textbf{Project page:} \url{https://kaiser-75.github.io/when2talk/}
\end{center}

\section{Introduction}
\label{sec:introduction}
Highly automated driving has moved beyond research prototypes into rider-only services operating on public roads in the United States. A
recent peer-reviewed analysis of an SAE Level~4~\cite{sae2018taxonomy} ride-hailing service examined 7.14 million rider-only miles 
accumulated across Phoenix, San Francisco, and Los Angeles~\cite{kusano2024comparison}. At this level of automation, the system can 
perform the complete dynamic driving task within its operational design domain, shifting the human occupant from an active driver to a 
passenger~\cite{on2021taxonomy}. These deployments make the passenger experience inside highly automated vehicles an increasingly 
relevant interaction design concern. Passengers may monitor the roadway, use a phone, participate in video conferences, or interact with 
increasingly capable voice and AI assistants. A proactive in-cabin agent is a vehicle-integrated assistant that initiates spoken 
communication about ride events without an explicit passenger request. 

Prior work has shown that communication about vehicle actions can shape how passengers understand and evaluate automated driving, including 
their trust, comfort, reliance, confidence, and communication preferences~\cite{koo2015did,koo2016understanding,du2019look,forster2017increasing,10.1145/3610886,10.1145/3706598.3713088}. The value of
this communication depends on the situation in which it is
delivered~\cite{ha2020effects,10.1145/3610886,zhang2023impact}.  Passenger activity is particularly relevant because it shapes how
passengers allocate attention and the information they obtain from the roadway 
environment~\cite{10.1145/3313831.3376751,strauch2019real,hungund2023impact}. For example, a passenger monitoring the roadway during a 
pedestrian crossing may already see the cues associated with the vehicle's response, whereas those cues may be missed when attention is
directed toward a phone. Proactive in-cabin communication must therefore consider not only what information to provide, but also whether
and when it should be communicated.

This system-initiated communication places the proactive in-car agent within mixed-initiative interaction, in which either the user or the
system can initiate a contribution ~\cite{10.1145/302979.303030}. A proactive in-car agent can provide relevant ride information without 
waiting for a passenger request ~\cite{10.1145/3340631.3394840,kraus2021role}. Poorly timed communication can disrupt ongoing activities or 
reduce the passenger's sense of control ~\cite{10.1145/3706598.3714002,10.1145/3706598.3713357,10.1145/3706598.3714317}. Interruption 
research further shows that disruption depends on task state, delivery time, and the effort required to suspend and resume ongoing activity
~\cite{mcfarlane2002scope,10.1145/985692.985727,bailey2006need,10.1145/1357054.1357072}. Delivering information at task boundaries can 
reduce disruption~\cite{10.1145/1054972.1055100,10.1145/1879831.1879833}, while receptivity depends on both the information being 
communicated and the passenger's ongoing activity~\cite{10.1145/2858036.2858566}. 

Research on automotive communication further demonstrates that suitable communication moments cannot be determined from vehicle telemetry 
alone~\cite{10.1145/3290605.3300867}. Explanation timing can influence evaluations of automated 
driving~\cite{10.1145/3610886,du2019look,du2023cross}. Interruptibility varies across people and non-driving 
activities~\cite{10.1145/3744333.3747831}, and response delay affects the acceptability of automotive conversational 
agents~\cite{10.1145/3409120.3410651}. Together, these findings establish the importance of adapting communication timing to both the
driving situation and the passenger’s activity. Less attention has been given to whether a detected event needs to be verbalized at all, 
particularly when vehicle motion and the surrounding scene already provide cues about what is happening. Proactive in-cabin communication 
can therefore involve not only immediate delivery, but also deferring information to a later opportunity or leaving it unspoken.

We study this design problem through \sys{}, which implements a context-sensitive (\textit{CS}) in-cabin communication as a policy 
that selects \textit{Immediate}, \textit{Delayed}, or \textit{Silent} delivery based on event priority and passenger activity. We 
compare \textit{CS} with an event-triggered (\textit{\ET{}}) policy  that delivers the event message immediately at every predefined 
ride event. Forty-one participants experienced both policies across six prespecified combinations of event priority and passenger 
activity. To provide an immersive passenger-seat experience, we delivered CARLA-based rider experience through a Varjo XR-4 VR 
headset. Participants completed \textit{Road Monitoring}, \textit{Intermittent Phone Use}, or \textit{Sustained Phone Use}. Event-
level questionnaires captured participants' evaluations after each ride event, while synchronized protocol logs recorded event and 
communication timing. The headset's eye tracker measured event-related pupil responses during \textit{Road Monitoring}. We limited this
analysis to \textit{Road Monitoring} because participants continuously viewed the simulated roadway, providing a common
visual task for time-locking pupil responses to ride events. During the two phone-use activities, participants directed
their visual attention toward the physical smartphone and therefore did not share the same roadway-focused viewing
condition. Figure~\ref{fig:teaser} provides an overview of the study environment, communication policies, experimental workflow, and 
research objective.

Our findings show that context-sensitive communication can reduce perceived interruption while preserving the value of in-cabin 
communication. Delaying communication can retain useful information while moving it to a more appropriate moment, whereas leaving 
information unspoken can avoid unnecessary communication when passengers already have access to relevant cues. This work provides 
controlled evidence on when communication is appropriate following a ride event, characterizes \textit{Immediate}, \textit{Delayed}, and 
\textit{Silent} communication across passenger activities and ride events, and contributes a reusable CARLA-based passenger-seat VR testbed 
for studying context-sensitive in-vehicle communication (see project page).

\section{Related Work}
\label{sec:related}

\subsection{Explanations in Automated Vehicles}

Automated vehicles can communicate their actions, intentions, and perception of the environment to support occupants' understanding of vehicle 
behavior. Explanation content shapes how people interpret automation. Explanations of why a vehicle acts can improve understanding and trust 
relative to descriptions of the action alone~\cite{koo2015did}, and speech output can increase trust, usability, and anthropomorphic 
perceptions~\cite{forster2017increasing}. Responses also depend on the driving situation. Perceived risk, delivery time, cultural context, 
correctness, and driving difficulty have all shaped evaluations of automated-vehicle explanations 
~\cite{ha2020effects,du2019look,du2023cross,10.1145/3610886,10.1145/3706598.3713088}.

Trust and transparency are therefore recurring concerns in AV communication. Trust develops through experience with
automation behavior, system transparency, and an occupant's understanding of system capabilities
~\cite{choi2015investigating,ekman2017creating,kraus2020more}. Explanations can shape how people interpret critical
transitions~\cite{korber2018have}, while communicating uncertainty can support more calibrated judgments of system
capability~\cite{10.1145/2516540.2516554}. Reviews of explainable automated driving accordingly organize the design space
around what should be explained, to whom, when, and through which modality~\cite{atakishiyev2024incorporating}.
Engagement in non-driving activities further changes how explanation modalities are received~\cite{zhang2023impact}.

This literature shows that explanation design extends beyond message content. How information is received also depends
on the driving situation, the occupant, and the circumstances in which communication occurs.

\subsection{Passenger Activity, Visual Access, and Attention}

Higher levels of automation allow occupants to devote attention to activities unrelated to vehicle control. Non-driving
activities affect visual allocation, workload, engagement, and readiness to redirect attention toward the roadway
~\cite{jaussein2021non,hungund2023impact}. Automated travel therefore involves movement between driving-related
information and other passenger activities~\cite{janssen2019interrupted}.

Passengers may nevertheless continue to sample the roadway while engaged in other activities. Occupants can
self-interrupt non-driving tasks to inspect the road, and interface configuration can change this behavior
~\cite{10.1145/3313831.3376751}. Gaze toward safety-relevant areas also varies with the driving situation and automation mode
~\cite{strauch2019real}. Other studies have related gaze behavior to trust and monitoring of automated systems
~\cite{hergeth2016keep,walker2019gaze}. These findings show that passenger activity affects how much information an
occupant obtains directly from the driving environment.

Speech provides another channel for vehicle information without requiring visual-manual interaction, but spoken
interaction can introduce cognitive demand. In-vehicle voice systems have produced measurable distraction
~\cite{strayer2016talking,strayer2019assessing,loew2023impact}. Passenger activity therefore matters not only because
it affects receptivity to an incoming message, but also because it changes how much of the driving situation is already
available through direct observation.

\subsection{Timing and Selective In-Cabin Communication}

Prior work has examined when information can be delivered during a ride. Vehicle and contextual signals can help
identify suitable communication moments~\cite{10.1145/3290605.3300867}. Response delay affects the usability and acceptability
of automotive conversational agents~\cite{10.1145/3409120.3410651}, and explanation timing can change evaluations of highly
automated driving~\cite{10.1145/3610886}.

Interruption research provides a broader account of why the delivery moment matters. Interruption has been described as
a coordination problem between an ongoing activity and an incoming demand~\cite{mcfarlane2002scope}, with costs varying
across points in task execution~\cite{10.1145/985692.985727}. Interruptions delivered during an activity can increase completion
time, errors, annoyance, and anxiety relative to task boundaries~\cite{bailey2006need}. Context-aware delivery around
activity transitions can reduce perceived burden~\cite{10.1145/1054972.1055100}, while notification receptivity depends on task
structure and message characteristics~\cite{10.1145/1879831.1879833,10.1145/2858036.2858566}. Recent automotive work extends this logic
to non-driving activities by relating interruptibility to task content, mental state, gaze, and head pose
~\cite{10.1145/3744333.3747831}.

Prior work therefore establishes that explanation content, passenger activity, and communication timing jointly shape how in-vehicle 
information is received. Less is known about how these factors should be combined when a passenger-facing agent must decide whether a
detected ride event warrants immediate speech, can be communicated later, or need not be verbalized. \sys{} examines this decision at the
communication-policy level by comparing an event-triggered policy with a context-sensitive policy that selects among \textit{Immediate}, 
\textit{Delayed}, and \textit{Silent} communication while holding the canonical message constant whenever both policies speak.

\section{Research Questions}
\label{sec:rqs}

When a proactive in-car agent detects a ride event, the communication decision may depend on both the event and the
passenger's ongoing activity. We examine two policies across six prespecified combinations of event priority and
passenger activity.

\textbf{RQ1} How do Event \textbf{Priority}, Passenger \textbf{Activity}, and Communication \textbf{Policy} affect rider experience, 
including perceived communication appropriateness, interruption, trust, and usefulness?

\textbf{RQ2} How do \textbf{Priority}, \textbf{Activity}, and \textbf{Policy} interact in shaping passenger experience?

\textbf{RQ3} How do passengers perceive and reason about context-sensitive in-cabin communication?

\section{User Study}
\label{sec:study}

We conducted a controlled within-subject study comparing an event-triggered communication policy (\textit{ET}) with
a context-sensitive communication policy (\textit{CS}) during Level~4 automated-vehicle rides. The experiment followed a
$2\,(\textbf{Policy})\times2\,(\textbf{Priority})\times3\,(\textbf{Activity})$
repeated-measures design. The six \textbf{Priority $\times$ Activity} combinations defined the experimental contexts, and each
participant experienced both communication policies within every context, yielding 12 event-level observations per
participant. Each context was instantiated using matched scenario variants across the two policy blocks. The variants
differed in surface details such as actor appearance while preserving the event meaning, vehicle response, passenger
activity, canonical message, and corresponding \textit{\CS{}} action.

\subsection{Apparatus and Study Environment}
\label{sec:apparatus}

We built an immersive passenger-seat study environment using CARLA~0.9.15~\cite{dosovitskiy2017carla}. A virtual
front-passenger viewpoint was attached to the ego vehicle and presented through a Varjo XR-4 mixed-reality head-mounted
display (HMD). The headset's binocular eye tracker recorded gaze and pupil-diameter data at 200~Hz. Before the
experimental rides, participants completed Varjo's gaze-calibration procedure by fixating on the displayed targets. The
experimenter monitored a mirrored view of the passenger scene on an external display.

A mixed-reality pass-through region over the participant's lap showed their hands and physical smartphone while the
simulated roadway remained visible through the HMD. A synchronized smartphone application supported the \textit{Intermittent Phone Use}
and \textit{Sustained Phone Use} task conditions, displayed activity instructions, and collected event-level questionnaire responses. A
Python-based study controller coordinated the CARLA route, scripted events, passenger activities, communication behavior,
survey pauses, and protocol logging. The controller also synchronized CARLA event markers with the eye-tracking record
on a common experimental timeline.

Ride events were enacted through scripted actors and vehicle responses within the simulated scene. In the pedestrian
condition, for example, a pedestrian entered the roadway and the ego vehicle yielded while the participant observed the
event from the front passenger seat. This synchronized configuration allowed the ride event, passenger activity,
communication behavior, questionnaire, and eye-tracking record to be interpreted relative to the same event timeline.

\begin{figure*}[t]
  \centering
  \includegraphics[width=\textwidth]{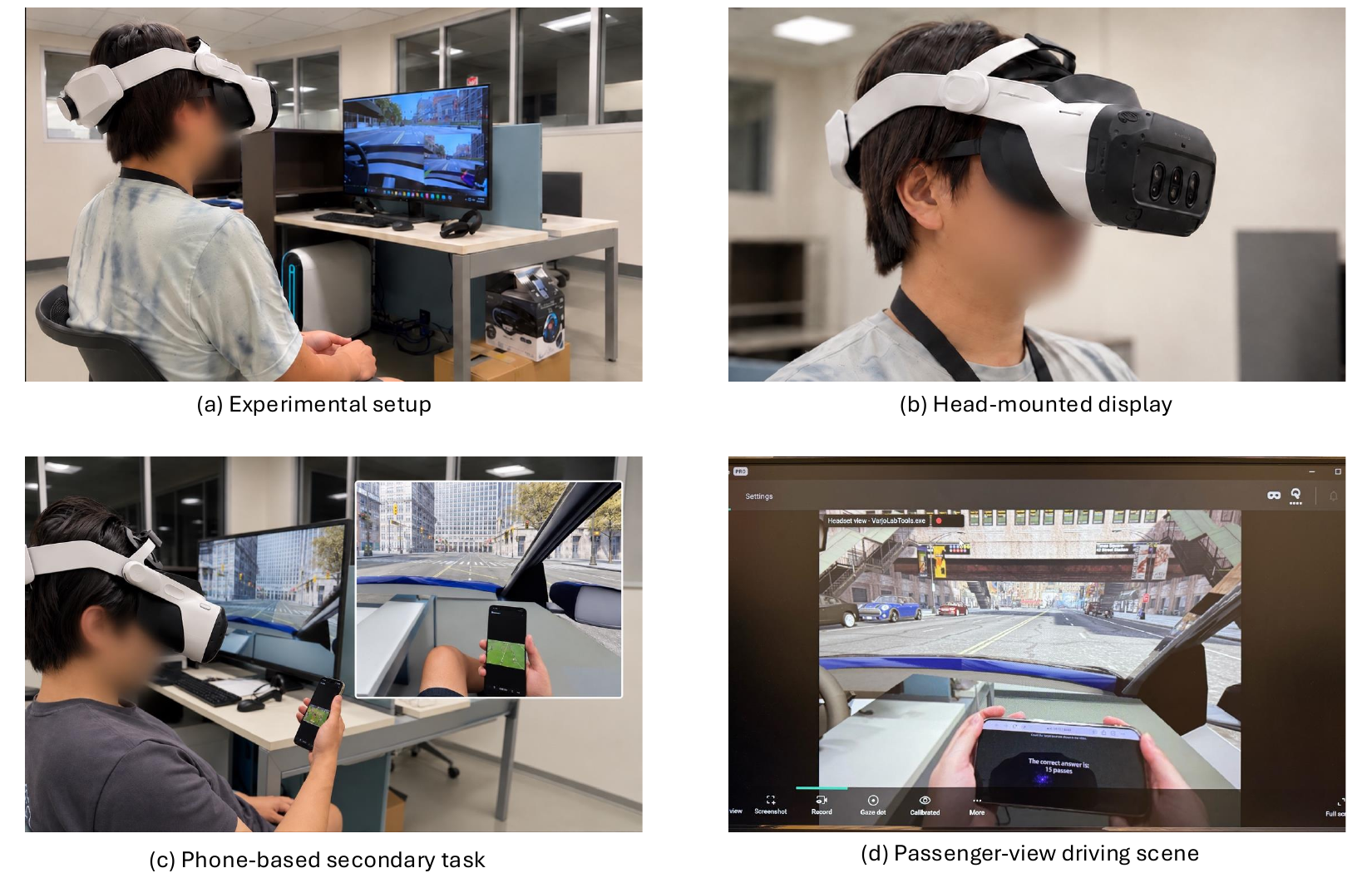}
  \caption{When2Talk experimental apparatus. Participants experienced scripted CARLA ride events from the front
   passenger seat through a Varjo HMD while a synchronized smartphone application controlled the assigned passenger
  activity.}
  \label{fig:setup}
\end{figure*}

The AV messages were prerecorded before participant trials. For a given ride event, the same canonical recording was
used whenever \textit{\ET{}} and \textit{\CS{} }both communicated, holding message wording and voice characteristics constant across
policies. Audio was delivered through headphones at a fixed study volume. Figure~\ref{fig:setup} shows the physical apparatus and the 
synchronized passenger-seat study environment.

\subsection{Independent Variables and Experimental Contexts}
\label{sec:design}

\begin{figure*}[t]
  \centering
  \includegraphics[width=0.8\textwidth]{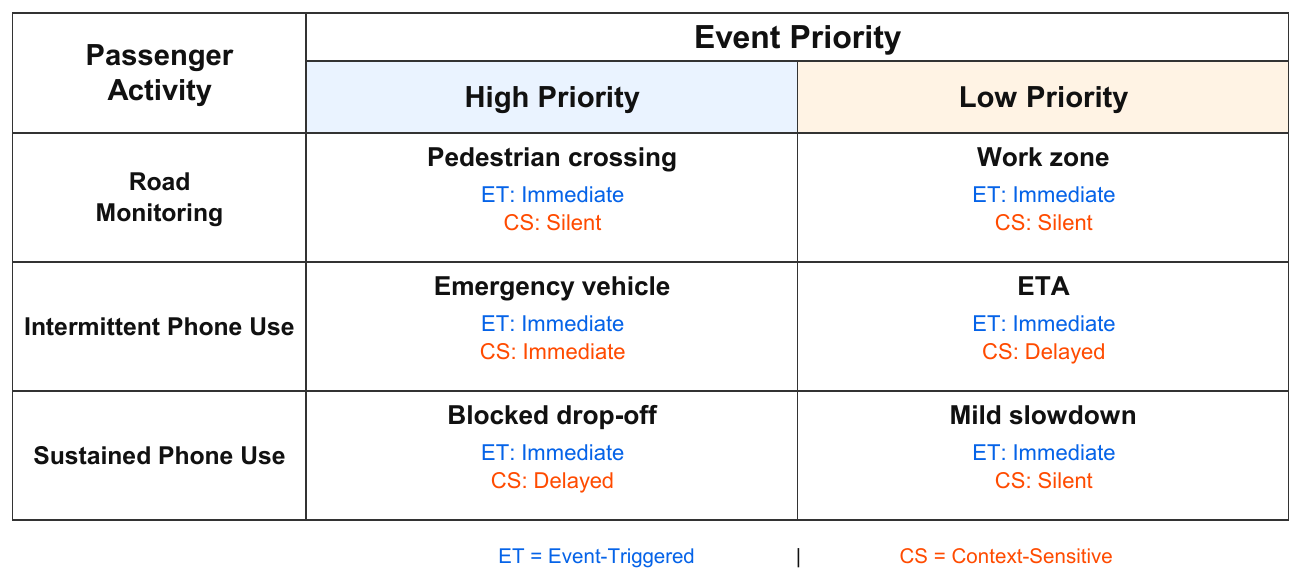}
  \caption{The $2\times2\times3$ within-subject design. Rows represent Passenger Activity, columns represent Event
  Priority, and each cell shows the prespecified ride-event context. Every cell was experienced under both \ET{} and
  \CS{}, yielding 12 event-level conditions per participant.}
  \Description{A factorial matrix showing three passenger activities by two event-priority levels, with the six
  ride-event scenarios and ET/CS communication actions.}
  \label{fig:design}
\end{figure*}

Table~\ref{tab:ivs} summarizes the three independent variables. We crossed Communication \textbf{Policy} (\ET{}, \CS{}), Event
\textbf{Priority} (\textit{High}, \textit{Low}), and Passenger \textbf{Activity} (\textit{Road Monitoring}, \textit{Intermittent Phone Use}, 
\textit{Sustained Phone Use}). \textbf{Activities} represented increasing intended costs of interruption. 
Each \textbf{Priority $\times$ Activity} combination defined one experimental context, and every participant experienced both \textbf{Policies} 
within each context. Figure~\ref{fig:design} summarizes the six contexts and corresponding \ET{} and \CS{} behavior.

\begin{table*}[t]
  \centering
  \caption{Independent variables, condition levels, and operational definitions used in the study.}
  \label{tab:ivs}
  \small
  \begin{tabular}{@{}p{0.20\textwidth}p{0.25\textwidth}p{0.49\textwidth}@{}}
    \toprule
    \textbf{Independent variable} &
    \textbf{Condition} &
    \textbf{Operational definition} \\
    \midrule

    Communication Policy
    & Event-Triggered (\ET{})
    & Delivers the fixed event message immediately at event onset. \\

    & Context-Sensitive (\CS{})
    & Uses the predefined \textit{Immediate}, \textit{Delayed}, or \textit{Silent} action assigned to the context. \\

    \addlinespace
    Event Priority
    & High
    & Passenger awareness or preparation was considered consequential within the scripted event. \\

    & Low
    & The event provided lower-consequence ride or route information within the study context. \\

    \addlinespace
    Passenger Activity
    & Road Monitoring (Low)
    & Participant monitors the forward roadway through the HMD without performing a secondary device activity. \\

    & Intermittent Phone Use (Medium)
    & Participant performs a smartphone counting task with periodic advertisement intervals that create prespecified task
    boundaries. \\

    & Sustained Phone Use (High)
    & Participant continuously watches standardized media content on the smartphone without prespecified task boundaries. \\

    \bottomrule
  \end{tabular}
\end{table*}

\subsubsection{Passenger Activities}
\label{sec:activities}

\textbf{Activity} was controlled through the synchronized smartphone application, with instructions presented at each
activity transition. The activities instantiated three attentional structures relevant to passenger travel. \textit{Road Monitoring} 
provided direct visual access to the driving environment. \textit{Intermittent Phone Use} introduced recurring task boundaries at which 
attention could shift, while \textit{Sustained Phone Use} maintained engagement with content away from the roadway. Prior studies of
automated driving have similarly examined visual monitoring, smartphone interaction, and sustained media viewing,
showing that attention allocation and interruptibility depend on the structure of the ongoing activity
~\cite{jaussein2021non,10.1145/3313831.3376751,hungund2023impact,10.1145/3744333.3747831}.

These structures also occur in other passenger activities. Browsing, messaging, and conversations may provide recurring
pauses at which communication can be introduced. Watching a movie, participating in a video conference, or engaging in
a focused conversation may hold attention for longer periods and provide fewer convenient communication moments. The
three conditions therefore provide controlled instances of attentional structures that can occur across a broader range
of passenger activities. Activity was analyzed as a categorical factor because the conditions differed in their interruption
structure and access to roadway information.

\subsubsection{Ride Events and Communication Policies}
\label{sec:policies}

\begin{figure*}[t]
  \centering
  \includegraphics[width=\textwidth]{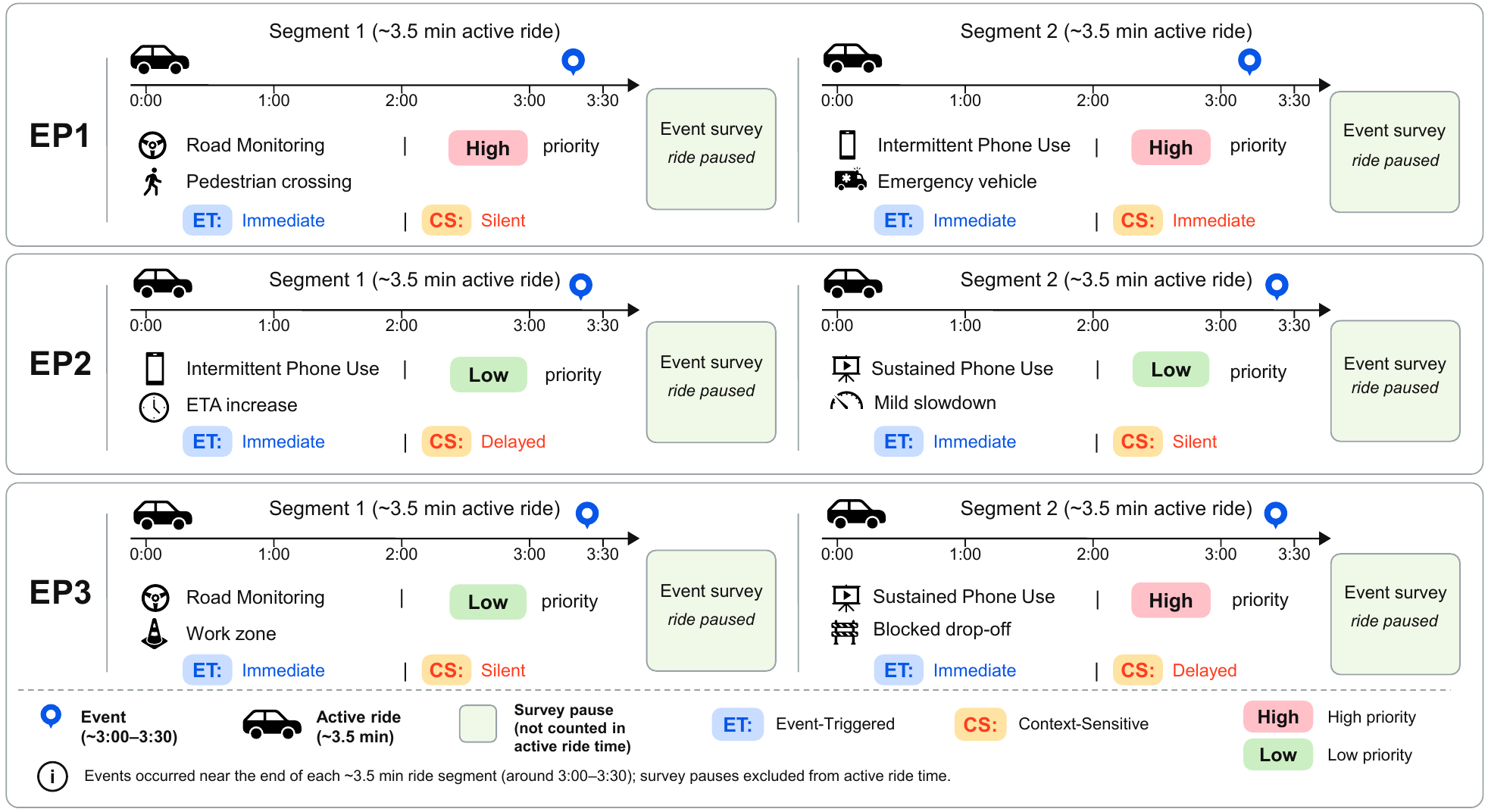}
  \caption{Episode and communication design. Each episode contained two scripted ride events separated by an event-level
  questionnaire and an activity transition. The figure shows the ride event and \ET{}/\CS{} action
  for each event. \textbf{Delayed} communication was tied to the prespecified later opportunity for the corresponding context.}
  \Description{Three episode timelines showing the two passenger activities, two ride events, event questionnaire, and
  ET/CS communication actions in each episode.}
  \label{fig:episodes}
\end{figure*}

Each \textbf{Priority $\times$ Activity} context was instantiated by one scripted ride event. Figure~\ref{fig:episodes}
summarizes the episode sequence and communication action used under each policy. Under \textit{\ET{}}, the agent delivered the canonical message 
immediately at event onset in every context. Under \textit{\CS{}},communication was predefined as \textit{Immediate}, \textit{Delayed}, or 
\textit{Silent}. \textit{Immediate} delivered the same message at event onset. \textit{Delayed} retained the message and delivered it at a 
prespecified opportunity 20 seconds later. For the \textit{ET} conditions, this opportunity coincided with a boundary in the \textit{Intermittent 
Phone Use} task. For the blocked-drop-off condition, it occurred at the corresponding later point during \textit{Sustained Phone Use}. 
\textit{Silent} produced no verbal message.

The six \CS{} actions were specified before participant recruitment. For each context, the researchers considered the information available from 
the roadway, and whether the event information would remain useful if delivered later. The researchers then finalized the \textit{Immediate}, 
\textit{Delayed}, or \textit{Silent} action shown in Figure~\ref{fig:design}. The policy was fixed throughout the study; no adaptive or learned 
model operated during participant trials.

\subsection{Procedure}
\label{sec:procedure}

The study protocol was approved by the University Institutional Review Board (IRB2025-31), and all participants provided informed consent before
beginning the study. Participants first completed a background questionnaire and received an introduction to the passenger-seat VR environment, 
the three Passenger Activities, the prerecorded in-cabin voice, and the rating scales. They then completed a familiarization ride before beginning
the experimental blocks. Each session lasted approximately one hour, and participants received \$20 for their participation.

Each participant completed two \textbf{Policy} blocks, one \textit{\ET{}} and one \textit{\CS{}}. Each block contained three ride episodes, with
two scripted events per episode, yielding six event-level observations per policy and 12 per participant. The three episode templates were 
\textit{Road Monitoring} $\rightarrow$ \textit{Intermittent Phone Use}, \textit{Intermittent Phone Use} $\rightarrow$ \textit{Sustained Phone 
Use}, and \textit{Road Monitoring} $\rightarrow$ \textit{Sustained Phone Use}.

Participants completed an event-level questionnaire after each ride event. After the first event in an episode, the ride
paused while the questionnaire was completed. The smartphone application then instructed the participant to transition
to the second activity before the ride resumed. The questionnaire following the second event was completed before the
next episode. Questionnaire time was not included in active ride time.

Policy order and scenario-set order were counterbalanced across four assignment groups, with group sizes of 10, 11, 10,
and 10. Twenty participants experienced \CS{} before \ET{}, and 21 experienced \ET{} before \CS{}. After each policy
block, participants completed the block-level questionnaire. After both blocks, participants answered seven open-ended
questions about when the vehicle should communicate immediately, defer a message, or remain silent.

\begin{figure}[t]
  \centering
  \includegraphics[width=0.6\linewidth]{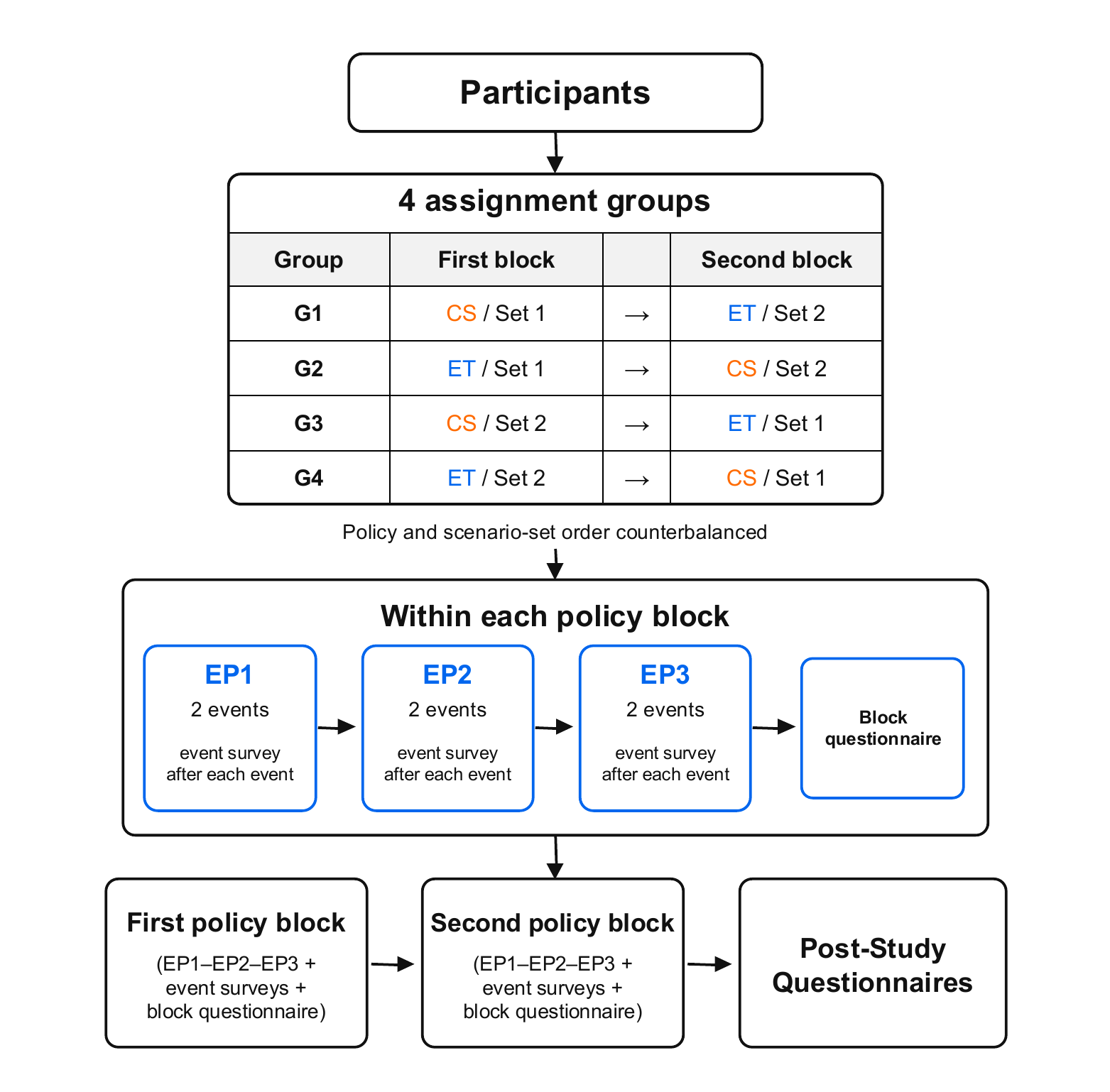}
  \caption{Counterbalancing and measurement schedule. Participants completed both policy blocks, with policy order 
  approximately balanced across the sample. Event-level questionnaires followed each ride event, block-level questionnaires 
  followed each three-episode policy block, and post-study questions followed completion of both blocks.}
  \Description{A study-flow diagram showing CS-first and ET-first counterbalancing, three episodes per block,
event-level questionnaires, block-level questionnaires, and post-study questions.}
  \label{fig:counterbalance}
\end{figure}

Figure~\ref{fig:counterbalance} summarizes the policy order, episode structure, and measurement schedule.

\subsection{Dependent Measures}
\label{sec:measures}

Table~\ref{tab:dvs} summarizes the dependent measures. After each ride event, participants rated communication
appropriateness, perceived interruption, trust in the communication decision, and usefulness using 7-point Likert
items. Communication appropriateness served as the primary event-level outcome. Usefulness was analyzed separately
because a message can provide valuable information even when its delivery is not judged to be the most appropriate
communication decision. The complete questionnaire wording and scale anchors are provided in the supplementary
material.

After each policy block, participants rated overall communication appropriateness, disruption, satisfaction, and three
trust items assessing confidence, reliability, and trustworthiness. The three trust items were averaged to form the
block-level trust composite. Baseline AV trust was collected before the experimental rides. 

For the two Road Monitoring events, the Varjo eye-tracking record provided an event-related pupil-response measure.
Pupil response represented the change in mean pupil diameter from the pre-event period to the corresponding scripted
event period.

\begin{table*}[t]
\centering
\caption{Dependent measures and measurement timing.}
\label{tab:dvs}
\small
\begin{tabular}{@{}lll@{}}
\toprule
\textbf{Measure} & \textbf{Timing} & \textbf{Scale} \\
\midrule
Communication appropriateness
& After each event & Single item, 1--7 \\
Perceived interruption
& After each event & Single item, 1--7 \\
Trust in communication decision
& After each event & Single item, 1--7 \\
Usefulness
& After each event & Single item, 1--7 \\
Event-related pupil response
& Road Monitoring events & Baseline-corrected diameter, mm \\
Overall communication appropriateness
& After each policy block & Single item, 1--7 \\
Perceived disruption
& After each policy block & Single item, 1--7 \\
Trust composite
& After each policy block & Three items, 1--7 \\
Satisfaction
& After each policy block & Single item, 1--7 \\
\bottomrule
\end{tabular}
\end{table*}

\subsection{Eye-Tracking and Pupil Response}
\label{sec:pupil-method}

The Varjo XR-4 headset recorded left- and right-eye pupil diameter in millimeters at 200~Hz. Eye-tracking samples were
synchronized with CARLA event markers through the Python study controller. Samples marked invalid by Varjo were excluded before aggregation.

For each participant, policy, and Road Monitoring event, mean pupil diameter was calculated over the 1~s immediately 
preceding event onset and over the corresponding scripted event, from onset to resolution. A 1~s pre-event baseline was
used following common pupillometry practice for event-related analyses~\cite{steinhauer2022publication}. Event-related
pupil response was calculated by subtracting the pre-event mean from the event-period mean.

\subsection{Data Analysis}
\label{sec:analysis}

Each event-level questionnaire outcome was analyzed using a
$2\,(\text{Communication \textbf{Policy}})\times2\,(\text{Event \textbf{Priority}})\times3\,(\text{Passenger \textbf{Activity}})$
repeated-measures ANOVA. We report $F$, $p$, and partial $\eta_p^2$ for all main effects and interactions.
Greenhouse--Geisser corrections were applied to effects involving Passenger Activity.

Significant interactions involving \textbf{Policy} were followed by paired comparisons of \ET{} and \CS{} within
the six  \textbf{Priority $\times$ Activity} contexts. Comparisons were Holm-corrected within each outcome, and
paired Cohen's $d_z$ was reported. Overall policy differences were calculated by averaging each participant's ratings
across Event Priority and Passenger Activity and were tested using paired $t$-tests with 95\% confidence intervals and
$d_z$.

Effect sizes and confidence intervals were calculated from \CS{}--\ET{} difference scores. Positive values indicate
higher ratings under \CS{}. Negative values favor \CS{} for perceived interruption and perceived disruption because
lower ratings indicate better outcomes.

Event-related pupil response was analyzed using a $2\,(\text{Policy})\times2\,(\text{Road-Monitoring Event})$ repeated-measures ANOVA. The two within-event
comparisons were Holm-corrected and reported with paired $d_z$.

Block-level outcomes were analyzed using paired $t$-tests with 95\% confidence intervals and $d_z$. Internal consistency
of the three-item trust composite was assessed using Cronbach's $\alpha$. Presentation-order effects were examined using
Welch independent-samples $t$-tests on participant-level \CS{}--\ET{} differences across the two policy-order groups.

Pearson correlations examined associations between baseline AV trust and participant-level \CS{}--\ET{} differences for
the four event-level outcomes. Correlations were calculated for the overall differences and separately for the
\textit{Immediate}, \textit{Delayed}, and \textit{Silent} contexts. For each action, ratings were averaged across its
corresponding contexts within each policy before calculating the difference scores. Holm correction was applied across
the four outcomes separately within each analysis.

\subsection{Post-Study Response Analysis}
\label{sec:poststudy-analysis}

To address RQ3, we reviewed responses to the seven open-ended post-study questions for recurring rationales concerning \textit{Immediate}, 
\textit{Delayed}, and \textit{Silent} communication. Responses were compared across participants and questions and grouped according to common 
considerations. The resulting response patterns concerned event consequence, passenger activity, the continuing value of deferred information, 
interpretations of silence, and confirmation needs. Representative quotations are identified by participant ID.

\subsection{Participants}
\label{sec:participants}

We recruited 41 participants through university mailing lists and a participant recruitment panel. All participants
completed the study and were included in the analysis. Participants received \$20 for completing the approximately
one-hour study.

The sample included 31 male and 10 female participants. Ages ranged from 19 to 30 years ($M=26.27$, $SD=2.64$). Most
participants reported no prior automated-vehicle experience (33 none, 8 moderate) and no prior virtual-reality experience
(37 none, 3 moderate, 1 extensive). Baseline trust in automated vehicles ranged from 1 to 6 on a 7-point scale
($M=3.51$, $SD=1.42$, median $=4$). Thirty-eight participants reported holding a valid U.S. driving license.

\section{Results}
\label{sec:results}

\subsection{Quantitative Results}
\label{sec:results-quant}

The quantitative analyses examined overall differences between \ET{} and \CS{} (RQ1) and whether these differences
varied across Event Priority and Passenger Activity (RQ2). We also analyzed event-related pupil response during Road
Monitoring, individual differences, and block-level evaluations.

\begin{figure*}[t]
  \centering
  \includegraphics[width=\textwidth]{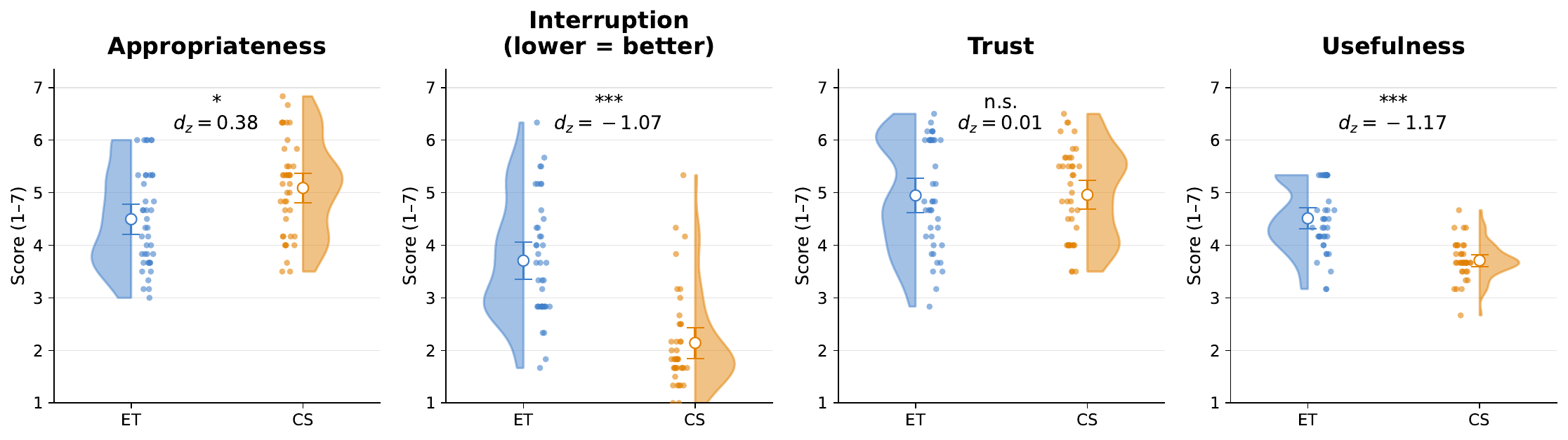}
  \caption{Overall event-level evaluations of \ET{} and \CS{} ($N=41$), averaged across Event Priority and Passenger 
  Activity. Half-violins and dots show the distributions of participant-level means; white circles and error bars indicate
  the grand mean and descriptive 95\% confidence interval. Lower scores indicate better outcomes only for perceived 
  interruption. Significance annotations report paired policy comparisons (* $p<.05$; *** $p<.001$; n.s. $p\geq.05$), and 
  $d_z$ denotes paired Cohen's $d_z$.}
  \Description{Four participant-level distribution plots compare ET and CS for appropriateness, perceived
  interruption, trust, and usefulness.}
  \label{fig:main-effects}
\end{figure*}

Figure~\ref{fig:main-effects} shows participant-level \textbf{Policy} scores averaged across \textbf{Priority} and \textbf{Activity.} 
Communication appropriateness was higher under \CS{} ($M=5.09$) than \ET{} ($M=4.50$), $t(40)=2.44$, $p=.019$,
95\% CI $[0.10,1.08]$, $d_z=.38$. Perceived interruption was lower under \CS{} ($M=2.14$) than \ET{}
($M=3.71$), $t(40)=-6.82$, $p<.001$, 95\% CI $[-2.03,-1.10]$, $d_z=-1.07$. Trust scores were
$M_{\ET}=4.95$ and $M_{\CS}=4.96$, $t(40)=0.05$, $p=.959$, 95\% CI $[-0.46,0.49]$, $d_z=.01$.
Usefulness was higher under \ET{} ($M=4.51$) than \CS{} ($M=3.71$), $t(40)=-7.49$, $p<.001$,
95\% CI $[-1.02,-0.58]$, $d_z=-1.17$.

\subsubsection{Communication Appropriateness (RQ1, RQ2)}
\label{sec:results-appropriateness}

Communication appropriateness was higher under \CS{} than under \ET{}, $F(1,40)=5.97$, $p=.019$, $\eta_p^2=.130$. The \textbf{Policy} effect
varied with \textbf{Priority}, $F(1,40)=14.84$, $p<.001$, $\eta_p^2=.271$, and \textbf{Activity}, $F(1.70,67.89)=17.29$, $p<.001$,
$\eta_p^2=.302$. The \textbf{Policy $\times$ Priority $\times$ Activity} interaction was not significant, $F(1.84,73.58)=1.05$, $p=.351$, 
$\eta_p^2=.025$.

Within-context comparisons showed higher appropriateness under \CS{} in the two \textit{Delayed} contexts, blocked
drop-off and ETA increase, and in the \textit{Silent} mild-slowdown context ($d_z=.51$--$.70$, all Holm-adjusted
$p\leq.010$). The remaining three comparisons did not reach significance after Holm correction. The overall
appropriateness difference was concentrated in these three contexts.

Appropriateness also varied with \textbf{Priority}, $F(1,40)=18.12$, $p<.001$, $\eta_p^2=.312$, and 
\textbf{Activity}, $F(1.63,65.35)=30.26$, $p<.001$, $\eta_p^2=.431$. Averaged across Policy and Passenger Activity,
appropriateness was higher in High-Priority contexts ($M=5.05$) than in Low-Priority contexts ($M=4.54$). Averaged
across \textbf{Policy} and \textbf{Priority}, ratings were highest during \textit{Intermittent Phone Use} ($M=5.37$), followed by \textit{Road
Monitoring} ($M=4.52$) and \textit{Sustained Phone Use} ($M=4.48$). The \textbf{Priority $\times$ Activity} interaction was significant,
$F(1.64,65.44)=23.09$, $p<.001$, $\eta_p^2=.366$.

\subsubsection{Perceived Interruption (RQ1, RQ2)}
\label{sec:results-interruption}

A significant main effect of \textbf{Policy} was observed, $F(1,40)=46.56$, $p<.001$, $\eta_p^2=.538$. The \textbf{Policy }effect varied with 
\textbf{Priority}, $F(1,40)=22.82$, $p<.001$, $\eta_p^2=.363$, and \textbf{Activity},
$F(1.97,78.77)=31.91$, $p<.001$, $\eta_p^2=.444$. The \textbf{Policy $\times$ Priority $\times$ Activity} interaction
was also significant, $F(1.93,77.14)=12.53$, $p<.001$, $\eta_p^2=.239$.

\textit{\CS{}} received lower interruption ratings in all five contexts in which it delayed or withheld speech
($|d_z|=.73$--$1.11$, all Holm-adjusted $p<.001$). In the \textit{Immediate} emergency-vehicle context, where
the policies used the same message and delivery timing, interruption was slightly lower under \textit{\ET{}}
($M_{\ET}=1.90$, $M_{\CS}=2.46$), $d_z=.34$, Holm-adjusted $p=.036$.

A significant main effect of \textbf{Priority} ($F(1,40)=20.53$, $p<.001$, $\eta_p^2=.339$) and
\textbf{Activity} ($F(1.84,73.45)=21.96$, $p<.001$, $\eta_p^2=.354$) was also observed. Averaged across \textbf{Policy} and 
\textbf{Activity}, interruption was lower in High-Priority contexts ($M=2.73$) than in Low-Priority contexts ($M=3.12$).
Averaged across \textbf{Policy} and \textbf{Priority}, interruption was highest during \textit{Sustained Phone Use} ($M=3.36$), followed
by \textit{Intermittent Phone Use} ($M=2.74$) and Road Monitoring ($M=2.68$). The \textbf{Priority $\times$ Activity} interaction was
significant, $F(1.98,79.11)=23.26$, $p<.001$, $\eta_p^2=.368$.

\subsubsection{Trust (RQ1, RQ2)}
\label{sec:results-trust}

The main effect of \textbf{Policy}  was not significant, $F(1,40)=0.00$, $p=.959$, $\eta_p^2=.000$. The Policy effect
varied significantly, nevertheless,  with \textbf{Priority}, $F(1,40)=8.81$, $p=.005$, $\eta_p^2=.180$, and \textbf{Activity},
$F(1.85,74.01)=6.93$, $p=.002$, $\eta_p^2=.148$. The \textbf{Policy $\times$ Priority $\times$ Activity} interaction
was not significant, $F(1.88,75.01)=0.50$, $p=.595$, $\eta_p^2=.012$.

Trust varied with \textbf{Priority}, $F(1,40)=27.11$, $p<.001$, $\eta_p^2=.404$, and \textbf{Activity}, $F(1.83,73.37)=5.29$, $p=.009$, 
$\eta_p^2=.117$. Averaged across \textbf{Policy} and \textbf{Activity}, trust was higher in High-Priority contexts ($M=5.18$) than in 
Low-Priority contexts ($M=4.72$). Averaged across \textbf{Policy} and \textbf{Priority}, trust was highest during \textit{Intermittent Phone Use} 
($M=5.19$), followed by Road Monitoring ($M=4.92$) and \textit{Sustained Phone Use} ($M=4.75$). The Priority $\times$ Activity interaction was 
significant, $F(1.98,79.31)=11.52$, $p<.001$, $\eta_p^2=.224$.

\subsubsection{Perceived Usefulness (RQ1, RQ2)}
\label{sec:results-usefulness}

Usefulness showed a significant \textbf{Policy }effect favoring \ET{}, $F(1,40)=56.14$, $p<.001$, $\eta_p^2=.584$. In addition, the 
\textbf{Policy} effect varied with \textbf{Activity}, $F(1.99,79.78)=45.79$, $p<.001$, $\eta_p^2=.534$, but not with 
\textbf{Priority}, $F(1,40)=0.82$, $p=.370$, $\eta_p^2=.020$. The \textbf{Policy $\times$ Priority $\times$ Activity} interaction
was significant, $F(2.00,79.98)=11.95$, $p<.001$, $\eta_p^2=.230$.

\textit{\ET{}} received higher usefulness ratings in the three \textit{Silent} contexts
($|d_z|=.69$--$1.63$, all Holm-adjusted $p<.001$). No significant usefulness differences were detected in the two
\textit{Delayed} contexts or the \textit{Immediate} context.

Usefulness also varied with \textbf{Priority}, $F(1,40)=408.82$, $p<.001$, $\eta_p^2=.911$, and \textbf{Activity},
$F(1.78,71.33)=182.65$, $p<.001$, $\eta_p^2=.820$. Averaged across \textbf{Policy} and \textbf{ Activity}, usefulness
was higher in High-Priority contexts ($M=4.80$) than in Low-Priority contexts ($M=3.43$). Averaged across \textbf{Policy} and
\textbf{Priority}, usefulness was highest during \textit{Intermittent Phone Use} ($M=5.00$), followed by \textit{Sustained Phone Use}
($M=3.97$) and Road Monitoring ($M=3.37$). The \textbf{Priority $\times$ Activity} interaction was significant,
$F(1.96,78.33)=42.31$, $p<.001$, $\eta_p^2=.514$.

Across outcomes, the within-context comparisons showed that that \textbf{Policy} differences varied across the six
tested contexts (Table~\ref{tab:context-comparisons}).
\begin{table*}[t]
  \centering
  \caption{Paired comparisons between \ET{} and \CS{} within the six tested contexts. Effect sizes were calculated from
  \CS{}--\ET{} difference scores; positive $d_z$ values indicate higher ratings under \CS{}. $p$-values are Holm-corrected
  across the six comparisons within each outcome. Statistically significant adjusted $p$-values are shown in bold. Lower ratings indicate better outcomes only for perceived interruption.}
  \label{tab:context-comparisons}
  \scriptsize
  \setlength{\tabcolsep}{5.5pt}
  \begin{tabular}{@{}lllrrrr@{}}
    \toprule
    \textbf{Outcome} & \textbf{Event context} & \textbf{\CS{} action} &
    $M_{\ET}$ & $M_{\CS}$ & $d_z$ & $p$ \\
    \midrule
    Appropriateness & Pedestrian & Silent & 4.66 & 4.46 & -0.07 & 1.000 \\
     & Emergency vehicle & Immediate & 6.05 & 6.07 & 0.09 & 1.000 \\
     & Blocked drop-off & Delayed & 4.05 & 5.00 & 0.51 & \textbf{.010} \\
     & Work zone & Silent & 4.37 & 4.61 & 0.12 & 1.000 \\
     & ETA increase & Delayed & 4.17 & 5.20 & 0.52 & \textbf{.009} \\
     & Mild slowdown & Silent & 3.68 & 5.20 & 0.70 & $\mathbf{<.001}$ \\
    \addlinespace
    Interruption & Pedestrian & Silent & 3.68 & 1.49 & -1.05 & $\mathbf{<.001}$ \\
     & Emergency vehicle & Immediate & 1.90 & 2.46 & 0.34 & \textbf{.036} \\
     & Blocked drop-off & Delayed & 4.34 & 2.51 & -1.08 & $\mathbf{<.001}$ \\
     & Work zone & Silent & 3.90 & 1.63 & -1.11 & $\mathbf{<.001}$ \\
     & ETA increase & Delayed & 4.02 & 2.56 & -0.73 & $\mathbf{<.001}$ \\
     & Mild slowdown & Silent & 4.39 & 2.20 & -0.99 & $\mathbf{<.001}$ \\
    \addlinespace
    Trust & Pedestrian & Silent & 5.29 & 4.63 & -0.29 & .392 \\
     & Emergency vehicle & Immediate & 5.88 & 5.56 & -0.26 & .431 \\
     & Blocked drop-off & Delayed & 4.73 & 5.00 & 0.13 & .637 \\
     & Work zone & Silent & 5.05 & 4.71 & -0.16 & .637 \\
     & ETA increase & Delayed & 4.44 & 4.88 & 0.20 & .607 \\
     & Mild slowdown & Silent & 4.29 & 4.98 & 0.30 & .392 \\
    \addlinespace
    Usefulness & Pedestrian & Silent & 4.71 & 2.39 & -1.63 & $\mathbf{<.001}$ \\
     & Emergency vehicle & Immediate & 5.88 & 5.93 & 0.05 & .756 \\
     & Blocked drop-off & Delayed & 5.10 & 4.78 & -0.26 & .306 \\
     & Work zone & Silent & 3.85 & 2.51 & -1.05 & $\mathbf{<.001}$ \\
     & ETA increase & Delayed & 3.98 & 4.22 & 0.21 & .385 \\
     & Mild slowdown & Silent & 3.56 & 2.44 & -0.69 & $\mathbf{<.001}$ \\
    \bottomrule
  \end{tabular}
\end{table*}

\subsubsection{Effects of Event Priority and Passenger Activity}
\label{sec}

\begin{figure*}[t]
\centering
\includegraphics[width=\textwidth]{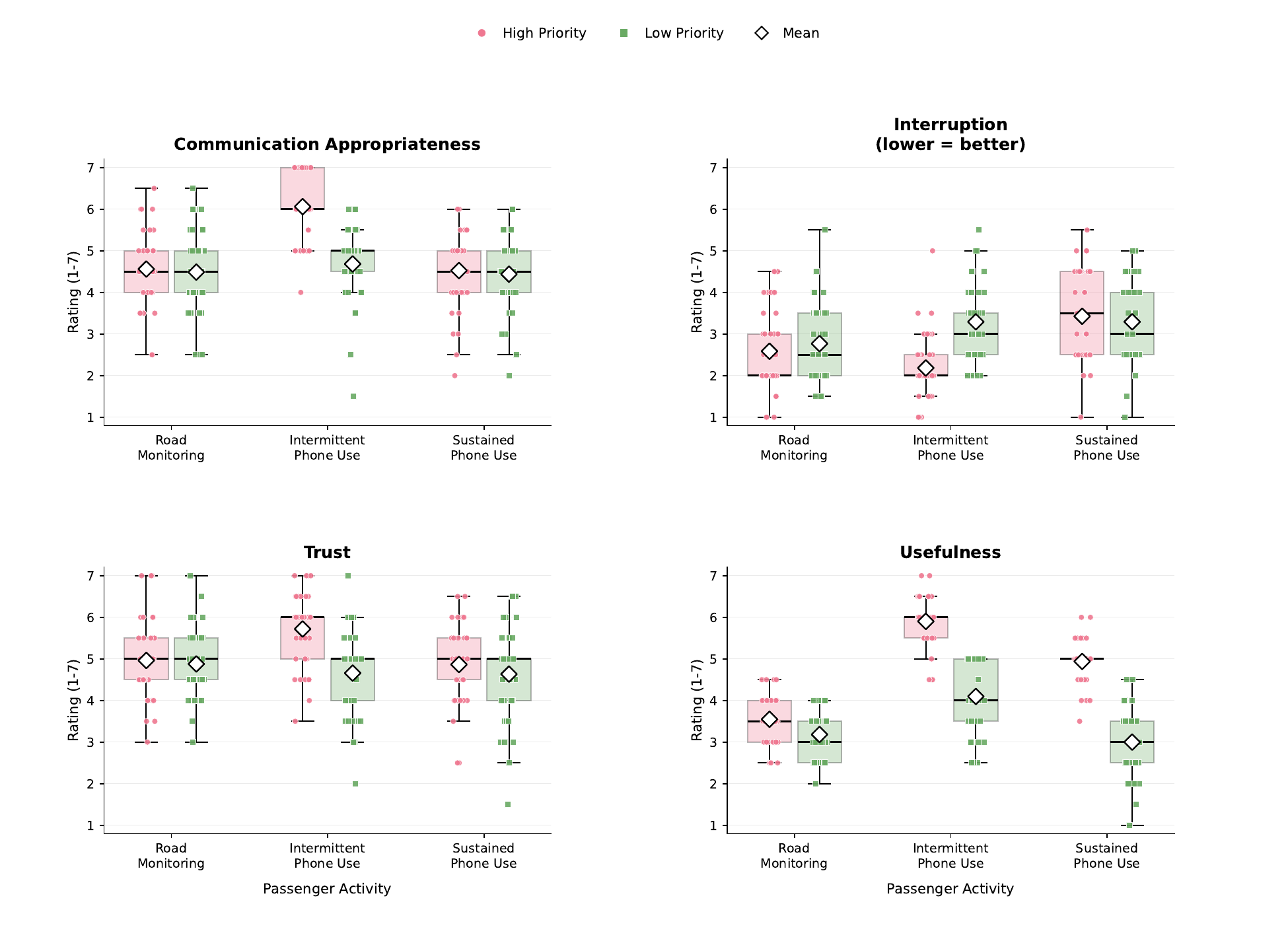}
\caption{Event-level outcomes across Event Priority and Passenger Activity, averaged across Communication Policy.
Each point represents one participant's mean across \ET{} and \CS{} within the corresponding context. Boxes show the
interquartile range, horizontal lines indicate medians, and whiskers extend to observations within 1.5 times the
interquartile range. White diamonds indicate means. Lower ratings indicate better outcomes only for perceived
interruption ($N=41$).}
\Description{Four box-and-point plots show communication appropriateness, perceived interruption, trust, and
usefulness across Road Monitoring, \textit{Intermittent Phone Use}, and \textit{Sustained Phone Use}. High- and Low-Priority
distributions are shown side by side. Each distribution contains participant-level ratings averaged across the two
Communication Policies.}
\label{fig:priority-activity}
\end{figure*}

Figure~\ref{fig:priority-activity} shows participant-level outcome distributions across Event Priority and Passenger
Activity after averaging each participant's ratings across Communication Policy. Across all four outcomes, Passenger
Activity patterns differed between High- and Low-Priority contexts, consistent with the significant
\textbf{Priority} $\times$ \textbf{Activity} interactions. Descriptively, the High- versus Low-Priority difference was largest during
\textit{Intermittent Phone Use} for appropriateness, perceived interruption, and trust. For usefulness, this difference was
pronounced during both phone-use activities.

\subsubsection{Pupil Response During Road Monitoring}
\label{sec:results-pupil}

Event-related pupil response was greater under \CS{} across the two Road Monitoring events. The
$2\,(\text{Policy})\times2\,(\text{Road-Monitoring Event})$
analysis showed a Policy main effect, $F(1,40)=17.47$, $p<.001$,
$\eta_p^2=.304$, and an Event main effect, $F(1,40)=6.18$, $p=.017$,
$\eta_p^2=.134$. The Policy $\times$ Event interaction was not significant,
$F(1,40)=2.50$, $p=.122$, $\eta_p^2=.059$.

During the high-priority pedestrian event, pupil response was greater under \CS{} ($M=0.16$, $SD=0.50$~mm) than
\ET{} ($M=-0.06$, $SD=0.27$~mm), $d_z=.50$, Holm-adjusted $p=.005$. During the low-priority work-zone event,
pupil response was also greater under \CS{} ($M=0.26$, $SD=0.38$~mm) than \ET{} ($M=0.17$, $SD=0.28$~mm),
$d_z=.38$, Holm-adjusted $p=.021$. In both Road Monitoring contexts, \CS{} withheld speech and \ET{} delivered
the canonical event message immediately.

\subsubsection{Individual Differences}
\label{sec:results-individual}

Baseline trust in automated vehicles was associated with participants' evaluations of \CS{} relative to \ET{}. Higher baseline trust 
corresponded to more positive overall \CS{}--\ET{} differences for appropriateness ($r=.79$), trust ($r=.72$), and usefulness ($r=.64$, all 
Holm-adjusted $p<.001$). The association with perceived interruption did not reach statistical significance ($r=-.14$, Holm-adjusted $p=.376$).

Action-specific analyses showed similar associations in the \textit{Silent} and \textit{Delayed} contexts. In the \textit{Silent} contexts,
higher baseline trust corresponded to more positive \CS{}--\ET{} differences for appropriateness ($r=.78$), trust ($r=.79$), and usefulness 
($r=.71$, all Holm-adjusted $p<.001$). The association with perceived interruption was not significant ($r=-.14$, Holm-adjusted $p=.394$). In 
the \textit{Delayed} contexts, baseline trust was associated with the differences in appropriateness ($r=.71$, Holm-adjusted $p<.001$) and trust
($r=.59$, Holm-adjusted $p<.001$), but not perceived interruption ($r=-.27$, Holm-adjusted $p=.173$) or usefulness ($r=.07$, Holm-adjusted 
$p=.648$). No association reached significance in the \textit{Immediate} context (all Holm-adjusted $p\geq.672$). Thus, baseline trust was most 
consistently associated with evaluations of actions that departed from immediate communication. No association with perceived interruption 
reached significance in the overall or action-specific analyses.

We did not observe an association between age and the overall Policy differences ($|r|=.03$--$.16$, all Holm-adjusted $p=1.000$). Overall 
Policy-difference scores also did not differ significantly by gender (all $p\geq.239$) or prior automated-vehicle experience (all $p\geq.251$).

\subsubsection{Block-Level Evaluation}
\label{sec:results-system}

At the block level, \CS{} received lower perceived-disruption ratings than \ET{} ($M_{\ET}=3.98$, $M_{\CS}=2.02$),
$t(40)=-10.38$, $p<.001$, 95\% CI $[-2.33,-1.57]$, $d_z=-1.62$.

No significant Policy difference was detected for block-level appropriateness ($M_{\ET}=4.59$, $M_{\CS}=5.05$),
$t(40)=1.34$, $p=.188$, 95\% CI $[-0.24,1.16]$, $d_z=.21$. The corresponding differences in block-level trust
($M_{\ET}=4.66$, $M_{\CS}=5.08$), $t(40)=1.37$, $p=.177$, 95\% CI $[-0.20,1.05]$, $d_z=.21$, and satisfaction
($M_{\ET}=4.71$, $M_{\CS}=5.17$), $t(40)=1.33$, $p=.192$, 95\% CI $[-0.24,1.17]$, $d_z=.21$, also did not reach
significance. The three-item trust composite showed high internal consistency under both policies ($\alpha=.97$ for
\ET{} and \CS{}).

No significant presentation-order differences were detected for any of the four event-level or four block-level
\CS{}--\ET{} difference scores (all $p\geq.435$).

\subsection{Post-Study Interviews (RQ3)}

The post-study responses provide further insight into how participants expected proactive in-car communication to adapt
across situations. Across the seven open-ended questions, five recurring themes described when communication should be 
\textit{Immediate}, \textit{Delayed}, or \textit{Silent}.

\subsubsection{Consequential events prompted a desire for immediate acknowledgment}
Participants expected immediate communication when an event affected their understanding of the ride, required
preparation, or explained a change in vehicle behavior. P005 wanted the vehicle to speak immediately about events such
as the emergency vehicle because hearing the message \textit{``confirms that the car noticed the event too.''} P029
similarly stated, \textit{``I still want high-priority information immediately even when I am busy.''} Participants also
emphasized that immediate communication could remain concise. P002 explained that \textit{``a short acknowledgment is
enough for me.''} These responses emphasized timely confirmation and concise delivery for consequential events.

\subsubsection{Passenger activity shaped timing preferences and perceived information needs}
Participants evaluated communication timing in relation to their ongoing activity. P004 explained that \textit{``the
passenger activity should influence the timing of the message.''} During visually demanding or engaging activities,
participants wanted routine information to be delivered more selectively. Road Monitoring created a different
information environment because passengers could observe parts of the driving situation directly. P020 preferred
\textit{``fewer messages when I am already visually monitoring the event.''} Activity also influenced tolerance for
communication during less active situations. P039 noted that \textit{``resting changes my tolerance for small
notifications more than road monitoring does.''} Passenger activity therefore shaped both receptivity to speech and the
amount of information already available from the surrounding scene.

\subsubsection{Deferral was acceptable when information remained useful}
Participants described \textit{Delayed} communication as appropriate for information that remained useful and could be delivered
at a suitable later moment. Participants repeatedly identified pauses and task boundaries as appropriate delivery opportunities. P031
reported that \textit{``the best moments were when the message arrived after my task naturally paused.''} Several
responses also emphasized that deferred information should remain available. P039 was \textit{``comfortable with delay,
but not with losing the update completely.''} P040 similarly explained that \textit{``a delayed message works for me if
the information is still useful at the later moment.''}

This theme aligns with the quantitative pattern in the \textit{Delayed} contexts, where \CS{} received higher appropriateness and
lower interruption ratings while usefulness did not differ significantly between policies.

\subsubsection{Silence reduced narration but removed explicit confirmation}
Participants interpreted silence through two considerations. Some treated silence as appropriate restraint when an event
was already visible, low priority, or added little beyond what they had inferred. P009 explained, \textit{``I do not need
narration for things that are already obvious from the windshield.''} P020 likewise valued \textit{``quiet periods when
nothing required my attention.''} For these participants, silence reduced unnecessary narration when speech provided
limited additional value.

Other participants used acknowledgment as evidence that the automation had perceived the event. P005 described the
\textit{Silent} contexts as the moments they \textit{``trusted the communication decision least''} because they were unsure
whether the automation had noticed the event for the right reason. P010 similarly explained that silence
\textit{``removes useful evidence that the system detected the event.''} Silence could therefore communicate appropriate
restraint while changing the confirmation available to the passenger.

\subsubsection{Passengers differed in their need for explicit confirmation}
Participants expressed different expectations for acknowledgment, communication detail, and overall message frequency.
P034 explained that \textit{``the overall amount of speech matters less to me than whether each message feels
justified.''} Participants who valued acknowledgment often accepted concise messages or delayed details. P008 proposed
to \textit{``acknowledge the event first, then adapt the amount of detail to urgency and activity.''} P015 similarly
recommended brief acknowledgment for important detections, delayed non-urgent details, and the option to
\textit{``choose a quieter setting.''} Other participants favored quiet operation when speech repeated information that
was already apparent.

Participants' responses indicate that expectations for communication timing depended on event consequence, passenger
activity, whether the information remained useful at a later moment, and the passenger's need for explicit confirmation.
These considerations shaped when participants considered \textit{Immediate}, \textit{Delayed}, and \textit{Silent} communication appropriate.

\section{Discussion and Implications}
This work examines how proactive in-car agents should decide whether and when to communicate during automated rides.
Through a controlled passenger-seat study, we found that \textit{\CS{}} increased communication appropriateness and substantially
reduced perceived interruption relative to \textit{\ET{}}. These effects varied by the communication action selected under
\textit{\CS{}}. In the \textit{Delayed} contexts, \textit{\CS{}} received higher appropriateness and lower interruption ratings, while no
significant usefulness differences were detected. In the \textit{Silent} contexts, \textit{\CS{}} reduced interruption, whereas
the explicit messages delivered under \textit{\ET{}} received higher usefulness ratings. Our findings identify event consequence,
passenger activity, continuing information value, and confirmation need as key considerations when selecting among
\textit{Immediate}, \textit{Delayed}, and \textit{Silent} communication. These findings inform the design of proactive
in-car agents that adapt communication delivery to the ride event and the passenger's ongoing activity.

\subsection{Context-Sensitive Communication Generally Reduced Interruption}
Reduced interruption was the most consistent benefit of \textit{\CS{}}. As shown in Figure~\ref{fig:main-effects}, \CS{} substantially
reduced perceived interruption overall. Interruption ratings were lower in all five contexts where it
delayed or withheld speech. Participants also rated the \textit{\CS{}} block as less disruptive, connecting the
event-level pattern to their broader experience of the policy.

This result builds on prior work showing that suitable communication moments depend on the driving situation and the
occupant's current activity~\cite{10.1145/3290605.3300867,10.1145/3610886,du2023cross}. Interruption research similarly
links disruption to delivery time, task state, and the effort required to suspend and resume an ongoing
activity~\cite{mcfarlane2002scope,10.1145/985692.985727,bailey2006need}. Our findings extend these principles from
selecting a suitable delivery moment to deciding whether a detected ride event should produce speech immediately, later,
or not at all.

The absence of a significant policy effect on perceived trust is also noteworthy but not unexpected. Prior research suggests that transparency and 
explanations can facilitate trust by providing users with additional information about an automated vehicle’s status, intentions, capabilities, or 
limitations~\cite{HA2020271}. Compared to \textit{ET}, \textit{CS} policy did not introduce additional information about the vehicle and its 
surroundings; rather, it altered the timing of existing information or withheld communication when that information was considered redundant.Importantly,
communication preferences also varied with participants’ preexisting trust in AVs. Trust research distinguishes such preexisting tendencies from trust 
that develops during interaction with a particular automated system~\cite{hoff2015trust}. Participants with lower dispositional trust towards AV showed a 
greater preference for being kept informed, and treats communication as a confirmation that the AV has detected and is responding to the situation. Thus, the appropriate level of communication may depend not only on event and activity context, but also on passengers’ prior trust in automation.

\paragraph{Design implication.}
Context-sensitive in-car communication should first assess whether immediate speech adds value given the event
consequence, passenger activity, and information already available from the ride. Passenger attention should inform
this decision, but it should not by itself determine whether communication is suppressed, particularly when the event
has greater safety or trip relevance.

\subsection{Delayed and Silent Communication Produced Different Trade-offs}

The distinction between \textit{Delayed} and \textit{Silent} communication shows that reducing interruption does not require removing
information. In the \textit{Delayed} contexts, \textit{\CS{}} improved appropriateness and reduced interruption while usefulness remained
comparable to \textit{\ET{}}. This pattern suggests that participants accepted deferral when the information remained relevant and
could be delivered at a more suitable moment, consistent with prior work showing that notification receptivity depends
on ongoing activity and that delivery near task boundaries can reduce disruption ~\cite{10.1145/1054972.1055100,10.1145/1879831.1879833,10.1145/2858036.2858566}. For proactive in-cabin agents, event
onset therefore need not determine message onset. When information retains value over time, it can remain pending until a
delivery opportunity that competes less with the passenger's current activity.

Silence produced a different trade-off because it removed both interruption and explicit information. Although \textit{\CS{}}
reduced interruption in all three Silent contexts, its appropriateness advantage appeared only for the mild slowdown.
This pattern suggests that withholding lower-priority information can be appropriate when immediate communication offers
limited additional value, but that visibility of an event alone does not justify silence. The absence of an
appropriateness advantage for the pedestrian and work-zone events, despite passengers' sustained attention to the
roadway, indicates that some events may still warrant explicit communication because of their consequence or the value
of confirming that the vehicle recognized them. This interpretation is consistent with prior work showing that perceived
alert appropriateness depends on the urgency of the underlying driving situation~\cite{large2019investigating}. Passenger
attention should therefore inform the communication decision rather than operate as a standalone rule for suppressing
speech~\cite{marshall2007alerts}.

The usefulness results further clarify this distinction. In the Silent contexts, \textit{\ET{}} retained an advantage because it
provided explicit information that \textit{\CS{}} withheld. Silence can therefore reduce interruption without making the omitted
message valueless. The design question is whether the added information or confirmation justifies occupying the current
interaction moment. By contrast, deferral preserves both the information and the opportunity to communicate it later.

The event-related pupil results support the role of the roadway scene as an information source during Road Monitoring.
Visible events continued to elicit measurable responses when \textit{\CS{}} withheld speech, indicating that passengers remained
responsive to the event even without narration. At the same time, the lack of significant block-level differences in
appropriateness, trust, and satisfaction suggests that the main benefit of \textit{\CS{}} was not a uniform improvement in every
evaluation, but a more selective reduction of interruption across specific contexts.

These findings assign different roles to deferral and silence. Deferral is appropriate when information remains useful
but the current moment is poorly suited for delivery. Silence is appropriate when speech adds limited value beyond the
event, vehicle response, and the passenger's existing understanding. A communication policy should therefore consider
both information lifetime and the value of explicit acknowledgment rather than treating \textit{Delayed} and \textit{Silent} behavior as
interchangeable forms of reduced communication.

\paragraph{Design implication.}
Delayed communication should be treated as a distinct scheduling action rather than as a weaker form of silence. When
information remains useful after event onset, the agent should retain it, identify a more suitable delivery opportunity,
and reassess whether the information is still relevant before speaking.

\subsection{Confirmation Needs Shaped the Value of Silence}

The post-study responses help explain why messages delivered under \textit{\ET{}} received higher usefulness ratings in the
\textit{Silent} contexts. Participants described spoken communication as providing event information and confirming that
the vehicle had detected or interpreted the event. Some participants considered the roadway scene and vehicle response
sufficient. Others valued explicit acknowledgment even when they could already infer what was happening.

Confirmation was particularly important for consequential events. Participants generally expected these events to be
acknowledged promptly, but several indicated that a short acknowledgment would be sufficient. Their responses
distinguished confirmation that the vehicle had recognized an event from a detailed description of the event.

Baseline trust further differentiated evaluations of selective communication. Participants with higher baseline trust
had more favorable \CS{}--\ET{} differences for appropriateness, trust, and usefulness overall. These associations were
especially consistent in the \textit{Silent} contexts and were also present for appropriateness and trust in the
\textit{Delayed} contexts. Baseline trust was not significantly associated with interruption differences. Participants
therefore differed in how they evaluated withheld or delayed communication even though \textit{\CS{}} reduced interruption across
the contexts in which it changed delivery.

Prior work has shown that trust in automation develops through experience, system behavior, and
transparency~\cite{choi2015investigating,ekman2017creating,kraus2020more}. Our findings connect these expectations to
communication delivery. Passengers may agree that a spoken message is interruptive while differing in how much
information or confirmation they expect the vehicle to provide.

The value of confirmation also depends on what the passenger can observe. Intermediate feedback can support users when
an in-cabin assistant's processing is difficult to observe~\cite{10.1145/3772318.3790997}. During an automated ride,
roadway events, vehicle motion, and the vehicle's response can already provide evidence of what the system has perceived.
Observable vehicle behavior and explicit communication can therefore jointly support passenger
understanding~\cite{axelsson2022modeling}.

\paragraph{Design implication.}
Acknowledgment should be separated from explanation detail. A brief acknowledgment can confirm that the vehicle has
recognized a consequential event without requiring a full explanation at that moment. Additional information can then
be delayed or omitted depending on what the passenger already knows and how much confirmation is needed.

The findings identify four considerations for selecting among \textit{Immediate}, \textit{Delayed}, and
\textit{Silent} communication. Event consequence informs urgency. Passenger Activity indicates availability and access
to roadway information. Continuing information value indicates whether later delivery remains useful. Confirmation need
indicates whether observable vehicle behavior should be accompanied by an explicit acknowledgment.

\subsection{Scope and Future Work}

The study evaluated one predefined \textit{\CS{}} mapping across six combinations of Event Priority and Passenger Activity. Each
context was paired with one \textit{\CS{}} action, allowing us to evaluate the complete policy used in the study. Future studies
can compare \textit{Immediate}, \textit{Delayed}, and \textit{Silent} communication within matched events and evaluate
alternative policy mappings.

The delayed messages were delivered at prespecified later opportunities. Further studies can compare different delay
lengths and delivery opportunities, including task boundaries detected from passenger activity. Measures of information
recall, task-resumption time, and requests for clarification can extend the event-level ratings used in this study.

The synchronized CARLA-based passenger-seat VR testbed aligned ride events, passenger activities, communication actions,
questionnaires, protocol logs, and pupil measurements on a common experimental timeline. The testbed and its
custom-built VR assets provide a reusable foundation for comparing communication policies across automated-driving
events and passenger activities. Field studies can extend this work to spontaneous activities, social interaction,
ambient noise, and less predictable event sequences.

Future studies can also compare spoken messages with brief acknowledgments and other communication modalities to
examine how timing, information delivery, and confirmation jointly shape the passenger experience.

\section{Conclusion}
\label{sec:conclusion}

This work examined how proactive in-car agents should decide whether and when to communicate during automated rides.
Compared with \textit{\ET{}, \CS{}} increased communication appropriateness and substantially reduced perceived interruption.
These effects varied by the communication action. In the \textit{Delayed} contexts, \textit{\CS{}} received higher
appropriateness and lower interruption ratings, while no significant usefulness differences were detected. In the
\textit{Silent} contexts, \textit{\CS{}} reduced interruption, whereas the explicit messages delivered under \textit{\ET{}} received
higher usefulness ratings and provided confirmation that some participants valued. Our findings identify event
consequence, passenger activity, continuing information value, and confirmation need as key considerations when
selecting among \textit{Immediate}, \textit{Delayed}, and \textit{Silent} communication. These findings inform the
design of proactive in-car agents that adapt communication delivery to the ride event and the passenger's ongoing
activity.

\bibliographystyle{ACM-Reference-Format}
\bibliography{references}
\clearpage
\appendix
\section{Survey}

This appendix documents the background questions, event-level and block-level
questionnaires, and open-ended post-study questions used in the study. Unless
otherwise stated, questionnaire items used a seven-point agreement scale from
\textit{1 = Strongly disagree} to \textit{7 = Strongly agree}.

\subsection{Demographics and Prior Experience}

Before the experimental rides, participants answered the following background
questions:

\begin{itemize}
    \item \textbf{Age.} What is your age? (Numeric response.)
    \item \textbf{Gender.} What is your gender? (Categorical response.)
    \item \textbf{Prior AV experience.} How would you describe your previous
    experience with automated vehicles? (None, Moderate, Extensive.)
    \item \textbf{Prior VR experience.} How would you describe your previous
    experience with virtual reality? (None, Moderate, Extensive.)
    \item \textbf{Baseline AV trust.} How much do you trust automated vehicles?
    (\textit{1 = Not at all} to \textit{7 = Completely}.)
    \item \textbf{U.S. driving license.} Do you hold a valid U.S. driving
    license? (Yes, No.)
\end{itemize}

\subsection{Questionnaires}

\subsubsection{Event-Level Questionnaire}

Participants completed the following four items after each ride event. The
``communication decision'' referred to the behavior experienced in that event,
including a decision to speak immediately, deliver the message later, or remain
silent.

\begin{enumerate}
    \item The vehicle's communication decision was appropriate for this event.
    \item The vehicle's communication unnecessarily interrupted what I was
    doing.
    \item I trusted the vehicle's communication decision for this event.
    \item The vehicle's communication was useful for understanding the event.
\end{enumerate}

All four items used the seven-point agreement scale. Lower scores indicated a
more favorable outcome only for perceived interruption.

\subsubsection{Block-Level Questionnaire}

Participants completed the following items after each three-episode policy
block:

\begin{enumerate}
    \item Overall, the vehicle communicated at appropriate times during this
    block.
    \item Overall, the vehicle's communication disrupted what I was doing
    during this block.
    \item I felt confident in the vehicle's communication decisions during
    this block.
    \item The vehicle's communication decisions during this block were
    reliable.
    \item The vehicle's communication decisions during this block were
    trustworthy.
    \item Overall, I was satisfied with the vehicle's communication during this
    block.
\end{enumerate}

All six items used the seven-point agreement scale. Lower scores indicated a
more favorable outcome only for perceived disruption. Confidence, reliability,
and trustworthiness were averaged to form the block-level trust composite.

\subsection{Post-Study Questions}

After completing both policy blocks, participants answered seven open-ended
questions:

\begin{enumerate}
    \item In what situations should the vehicle speak immediately?
    \item In what situations should the vehicle wait before speaking?
    \item In what situations should the vehicle remain silent?
    \item How should the passenger's current activity affect when the vehicle
    communicates?
    \item How should the importance of an event affect when the vehicle
    communicates?
    \item Were there any communication moments that you disliked? Please
    explain.
    \item If you could give the vehicle rules for when to communicate, what
    rules would you give it?
\end{enumerate}
\end{document}